\documentclass[12pt,a4paper]{article}
\usepackage{graphicx,amsmath,amsfonts}
\def \ba{\begin{eqnarray}}\def\ea{\end{eqnarray}}
\def\bc{\begin{center}}\def\ec{\end{center}}
\def\nn{\nonumber\\}
\title{\Large \bf Challenge of lepton pair production
in peripheral collisions of relativistic ions}
\author{\large \bf S.R.Gevorkyan\footnote{On leave of absence from
Yerevan Physics Institute} and A.~V.~Tarasov}
\begin{document}
\maketitle
\bc
Joint Institute for Nuclear Research, 141980 Dubna,
Russia
\ec
\begin{abstract}
 The new approach to the lepton pair production in the Coulomb
 field of two highly relativistic nuclei was developed.Solving the
operator equation for lepton scattering in arbitrary Coulomb
field, we obtain the amplitudes  for lepton scattering in the
Coulomb potential in terms of light cone variables.\\
Using the Watson expansion for the amplitude of lepton scattering
on two centers we propose prescription which allows one to
construct the amplitude for lepton pair production in the Coulomb
field of two highly relativistic ions.\\
We show that for the certain sums of finite terms of the Watson
series numerous  cancellations  lead to infrared stability of
amplitude.
\end{abstract}
\section{Introduction}
For many years it has been known that the yield of lepton pairs in
the collision of two charged relativistic particles grows
with energy as the third power of logarithm ~\cite{LL,R}.\\
The present interest in the process of lepton pair production off
the Coulomb fields of two highly relativistic ions with charge
numbers $Z_1$ and $Z_2$
 \ba
 Z_1+Z_2\to e^+e^-+Z_1+Z_2,
\ea is aroused  mainly by operation of heavy ion colliders as RHIC
( Lorentz factor $\gamma=\frac{E}{M}=100$) and LHC
($\gamma=3000$).At such energies the cross section of process (1)
becomes  huge (tens kilobarns at RHIC, hundreds kilobarns at LHC
energies) so that its  precise knowledge becomes a pressing~\cite{BB}.\\
In the last years a lot of work ~\cite{SW} - ~\cite{G4} has been
done on the problem of Coulomb corrections (CC) in process (1)
(multiple photon exchanges between produced lepton pair and the
Coulomb fields of colliding ions).For the collisions of heavy ions
the relevant parameter $Z\alpha$ ($\alpha=\frac{e^2}{4\pi}$ is the
fine structure constant) is not small (for instance, for lead
$Z\alpha\approx 0.6$),thus the contribution of CC can be essential.\\
Due to Lorentz contraction the Coulomb fields of colliding
ultrarelativistic ions can be considered as very thin discs.This
gives hope that at relativistic energies one would obtain the
relative simple  expression for the amplitude of process (1)
with regard Coulomb corrections.\\
The authors of ~\cite{BMac} solved the Dirac equation for the
electron propagating in the Coulomb field of two highly
relativistic nuclei and using obtained the amplitude of the
process (1).The result was striking: allowance for the Coulomb
correction leads to multiplication of the Born amplitude by phase
factor,i.e. CC have not any impact on the cross section.\\
This result was criticized in the works ~\cite{ISS,LM,AB}.In fact,
it is well known ~\cite{BM} that the contribution of Coulomb
corrections to the process of lepton pair photoproduction off the
Coulomb field is given by a series in fine structure constant and
there are no reasons for disappearance of CC in ion collisions.\\
The next step was made in the number of works ~\cite{G1,G2,G3,G4},
where the multiphoton exchanges between the produced pair and the
ions Coulomb fields were calculated in perturbation theory.Using
the  Sudakov technique authors computed  Feynman diagrams relevant
to the first terms of the perturbative expansion (up to the fifth
order in $\alpha$ ) and tried to conjecture the general structure
of higher order terms ~\cite{G4}.Their main results can be
formulated as follows :  \\
1)Taking in account the Coulomb corrections does not reduce to the
simple phase factor,as is argued  in ~\cite{BMac,ERG}\\
2) Any term in the perturbation expansion of the amplitude $M_m^n
\sim (Z_1\alpha)^m(Z_2\alpha)^n $ is infrared stable,i.e.does not
depend on infrared regularization parameter ( screening radius)
$\lambda$.\\
The last property was proved in ~\cite{G1,G4} only for the first
terms of the perturbative expansion $(m,n\leq 2)$.On the other
hand generalization  to higher order done in ~\cite{G4},leads to
terms depending on the regularization parameter $\lambda$,thus
the further work in this direction is necessary.\\
 In the present work we propose a new approach to this
issue.We construct the amplitude of process (1) in the form of the
Watson series,thus expressing it through the operators relevant to
the lepton scattering on separate centres.We solve the
corresponding equations in the limit of infinite energies and
obtain the amplitudes for lepton scattering off any external
field.The problem of the regularization is
investigated.Considering the specific sets of the Watson
expansion,we show that the relevant amplitudes are infrared
stable.The prescription is proposed which allows one to calculate
the cross section of process (1) with high precision. \\
The following notations are used in the paper: $e,m $ are the
electron charge and mass;$u(p\prime),v(p)$ are electron and
positron spinors; $\gamma_\mu$ are Dirac matrices and
$\gamma_{\pm}=\gamma_{0}\pm\gamma_z$.We use the light cone
definition of four  momenta and coordinates $k_{\pm}=k_{0}\pm
k_z$, $x_{\pm}=x_0 \pm x_z$. Throughout the paper the transverse
components of  momenta and coordinates are defined as two
dimensional vectors.For instance, $\vec b_j$ are the impact
parameters of ions whereas $\vec x_i,\vec k_i$ are transverse
coordinates and momenta of leptons.The index j=1,2 is reserved for
quantities attached to relevant ions $Z_1,Z_2$.
\section{Watson expansion }
 The amplitude of  lepton pair production in an arbitrary external
electromagnetic field created by two potentials $A_{\mu}=A_\mu
^{(1)}+A_\mu ^{(2)}$ is given by \ba M= \bar
u(p\prime)T(p\prime,-p)v(p) \ea The operator $T(p,p\prime)$ can be
cast in the form of an infinite Watson series ~\cite{W}, which
allows one to express the amplitude of particle scattering on two
centers through the scattering amplitudes of each of them
$T_j(p,p\prime); j=1,2$.In short notation the Watson series for
scattering on two centers reads
 \ba T& = & T_1  + T_2  - T_1\otimes G \otimes T_2  - T_2 \otimes G \otimes T_1 \nn
  &+& T_1\otimes G\otimes T_2 \otimes G \otimes T_1 + T_2 \otimes G \otimes T_1
\otimes G \otimes T_2 + \ldots
 \ea
 $T_j$ are the separate amplitudes of lepton scattering in the
Coulomb field of ions $Z_1$ and $Z_2$ and G(k) is the lepton
casual Green function.In Fig.1 we depicted the possible  exchanges
in lepton pair production in accordance with various terms of the
Watson expansion.\footnote{After substituting (3) in (2) the first
two terms in (3) are identically zero,therefore later on we begin
numbering from the third term in (3)} The thick lines attached to
ions $Z_1,Z_2$ represent the full set of photon
exchanges between the lepton (electron or positron) and the ion.\\
The amplitudes $T_j(p\prime,p)$ satisfy the well-known (see
e.g.~\cite{T}) operator equation
 \ba
 T_j&=&V_j - V_j \otimes G\otimes T_j\nn
 V_j(p,p\prime)&=& e\gamma_\mu A_{\mu}^j(p-p\prime)
\ea
 Equation (4) for single amplitude can be solved in the
 case of ultrarelativistic energies.At such energies due to
 Lorentz contraction the Coulomb field of nucleus looks
 like very thin disc for which the Coulomb potential in moving
 system takes a simple form.In this case the solution of  equation (4)
 is found and can be presented in the following form:
 \ba
 T_1(p,p\prime)&=&(2\pi )^2 \delta (p_+ - p\prime_+)[\theta(p_+)f_1^+
 (\vec p -\vec p\prime) - \theta(- p_+)f_1^-(\vec p - \vec
 p\prime)]\gamma_+\nn
 T_2(p,p\prime)&=&(2\pi )^2 \delta (p_- - p\prime_-)
 [\theta(p_-)f^+_2(\vec p -\vec p\prime) -
\theta(- p_-)f^-_2(\vec p - \vec p\prime)]\gamma_-\\
f^{\pm}_j(\vec q)&=&\frac{i}{2\pi}\int d^2x e^{i\vec q  \vec x }
 [1-S_j^{\pm}(\vec x,\vec b_j)];S_j^{\pm}(\vec x,\vec b_j)
 =\exp(\pm i\chi (\vec x,\vec b_j));\nn
 \chi_j(\vec x,\vec b_j)&=&e\int\limits_{ - \infty }^\infty
 \Phi_j(\sqrt{( \vec x  - \vec b_j )^2 + z^2})dz
 \ea
The Coulomb potentials of ions  $\Phi_j$ and the eikonal-type
amplitudes for electron and positron scattering depend on the
impact parameters of colliding ions $\vec b_1,\vec b_2$.On the
other hand, as a result of translation invariance,the square of
amplitude (2) depend only on the difference
$\vec b=\vec b_1-\vec b_2$.\\
We adopt such normalization of the amplitude that the cross
section of the process (1) is
 \ba d\sigma=\frac{1}{(2\pi)^{10}}\int
|M(b,p,p\prime)|^2d^2b
\frac{d^3p}{2p_0}\frac{d^3p\prime}{2p_0\prime} \ea Substituting
the expressions (5) for $T_j$ in the Watson expansion (3), one can
calculated the cross section of process (1).Every consequent term
in the Watson series begins with higher order in the parameter
$Z\alpha$, thus one can calculate the cross section with a
desirable precision.This is true for the screened Coulomb
potential, for instance in the case of interaction of relativistic
atoms.But heavy ion colliders deal with ions whose Coulomb fields
are unscreened and for which the problem of amplitude
regularization demands special consideration.
\section{The challenge of infrared stability }
The Coulomb  phases (6) in the case of the screened Coulomb
potential reads \ba
 \chi_j(b) = e\int \Phi_j (\sqrt{b^2+z^2})dz=2Z_j\alpha
K_0(b\lambda_j )\to -2Z_j\alpha (\ln(b\lambda_j )+C)
\ea
 Here $\lambda_j$ are the screening radii, which can differ
 for different ions. Substituting  expressions (5,6) in the Watson
 expansion (3) leads to the  products of $S^\pm$-matrix elements some
of which do not depend on the screening parameter,for instance
 \ba
S_j^+(\vec x)S_j^-(\vec x\prime)=exp\left(2iZ_j\alpha \ln
\frac{|\vec x\prime -\vec b_j|}{|\vec x-\vec b_j|}\right) \ea
Nevertheless the majority of these products  are oscillating
functions of  $\lambda_j$.On the other hand,our experience from
photoproduction ~\cite{BM} and perturbation theory ~\cite{G2,G4}
tell us that the amplitude of the process under consideration must
be infrared stable, so all oscillating products have to be
cancelled in the full amplitude.\\
 To follow these cancellations, consider firstly  the case where one
of the ions,for instance $Z_2$, is light so that one can expands
the amplitude in the parameter $Z_2\alpha$.In the general case
Watson series (3) is infinite and  there are no reasons to
truncate it. However it is automatically cut off if one considers
the finite number of exchanged photons attached to one of the
nuclei (with any number of exchanges with another nuclei)\\
 Denoting the transverse momenta of  leptons in intermediate
 states by $\vec k_i$ and the transverse momenta of exchanged photons
 by $\vec q_i$ (see Fig.1) we introduce the following notations:
 \ba
\Omega_j(\vec q,\vec q\prime)=\frac{1}{(2\pi)^2} \int d^2 x d^2
x\prime\exp\left(i\vec q\vec x + i\vec q\prime \vec
x\prime\right)\left(1-S_j^+(\vec x)S_j^-(\vec x\prime)\right )\nn
=f_j^+(\vec q) f_j^-(\vec q\prime)-2\pi i\delta(\vec q)
f_j^+(\vec q) -2\pi i\delta(\vec q\prime) f_j^-(\vec q\prime)\\
f_j^+= -\sum_{n=1}^{\infty}{\frac{i^{n+1}}{n!}f_j^{(n)}};
f_j^-=\sum_{n=1}^{\infty}{\frac{(-i)^{n+1}}{n!}f_j^{(n)}};
f_j^{(n)}=\frac{1}{2\pi}\int d^2x e^{i\vec q\vec x}\chi^n(\vec
b_2-\vec x)
 \ea
Note that expressions (11) are nothing but
expansion of amplitudes from (6).\\
 To obtain the sum of all terms from Watson expansion (3)  relevant
to the first order exchange in $Z_2\alpha$ and any exchanges with
$Z_1$,it's enough to calculate the terms which are linear in
$T_2$.These terms correspond to the first three diagrams of Fig.1,
with only replacement of the thick line attached to
the ion $Z_2$ by a single photon exchange.\\
Using the above expressions after a lengthy but well known
algebra, we get: \ba
 M^{(1)}&=&\sum_{n=1}^{\infty} M_n^{(1)}=\frac{i}{8\pi}\int
\frac{\bar u(p\prime) \gamma_+\nu_1\gamma_-\nu_2\gamma_+
v(p)}{\mu_1p_++\mu_2p\prime_+} f_2^{(1)}(q_2)\Omega_1(q_1,q_3) d^2
k _1 d^2 k_2\nn
\nu_i&=&m-\vec k_i\vec\gamma ;\mu_i=m^2+\vec k_i^2
 \ea This expression does not depend on the
regularization parameter $\lambda_1$ as a result of nontrivial
cancellations among the different terms of the Watson
series.Passing in this expression to the impact parameter
representation upon the relevant Fourier transformations it is
easy to show that it is in
accordance with the results obtained in ~\cite{G3,BKS}.\\
As a next example of the independence from $\lambda_2$ we consider
the set of terms from the Watson series corresponding to two
photons attached to the ion  $Z_2$ (any number of exchanged
photons with $Z_1$).This contribution is provided by the first
four  diagrams of Fig.1, with obvious replacement of a set of
photon exchanges attached to ion $Z_2$ by one and two photon
exchanges.The result of our calculations reads
 \ba
 M^{(2)}&=&\sum_{n=1}^{\infty} M_n^{(2)}=-\frac{i}{(4\pi)^2}\int \frac{\bar
u(p\prime) \gamma_+\nu_1\gamma_-\nu_2\gamma_+
v(p)}{\mu_1p_++\mu_2p\prime_+}
f_2^{(2)}(q_2)\Omega_1(q_1,q_3)\ln\left(\frac{\mu_1p_+}{\mu_2p\prime_+}\right)
d^2 k _1 d^2 k_2\nn&-& \frac{i}{(4\pi)^3}\int \frac{\bar
u(p\prime) \gamma_+\nu_1\gamma_-\nu_2\gamma_+\nu_3\gamma_-
v(p)}{\mu_1\mu_3+\mu_2p\prime_+p_-}
f_2^{(1)}(q_2)f_2^{(1)}(q_4)\Omega_1(q_1,q_3)\nn&\times&\left[\ln
\left(\frac{\mu_1\mu_3}{\mu_2p\prime_+p_-}\right)+i\pi\right]
d^2k_1 d^2 k_2d^2k_3-\frac{i}{(4\pi)^3}\int \frac{\bar u(p\prime)
\gamma_-\nu_1\gamma_+\nu_2\gamma_-\nu_3\gamma_+
v(p)}{\mu_1\mu_3+\mu_2p\prime_-p_+}\nn&\times&
f_2^{(1)}(q_1)f_2^{(1)}(q_3)\Omega_1(q_2,q_4)\left[\ln
\left(\frac{\mu_1\mu_3}{\mu_2p\prime_-p_+}\right)+i\pi\right]
d^2k_1 d^2 k_2d^2k_3 \ea As in the previous case this expression
does not depend on the regularization parameter $\lambda_2$.We do
not cite here the next sets of Watson terms corresponding to three
and four photons attached to the ion $Z_2$ in view of their
inconvenience, but we verified that they also do not depend on
the regularization parameter $\lambda_2$.\\
To  investigate the problem of regularization in the general case,
we consider the first six terms of Watson expansion (3) (diagrams
a-f in Fig.1).It can be shown that this set consists of the
infrared stable term $M_s$ and the term $M_u$ depending on
$\lambda_j$, i.e.
 \ba \sum_{n,m=1}^{\infty} M_n^m=M^s+M^u\ea
We calculated the infrared stable part $M_s$ with the following
result
 \ba
 M^s &=&\frac{i}{(4\pi)^3}\int d^2k_1 d^2k_2d^2k_3\bar u(p\prime)
 [\gamma_+\nu_1\gamma_-\nu_2\gamma_+\nu_3\gamma_-
\Omega_1(q_1,q_3)\Omega_2(q_2,q_4)\frac{\ln(\frac{\mu_1\mu_3}
{\mu_2p\prime_+p_-})+i\pi}{\mu_1\mu_3+\mu_2p\prime_+p_-}\nn
 &+&\gamma_-\nu_1\gamma_+\nu_2\gamma_-\nu_3\gamma_+\Omega_2(q_1,q_3)
 \Omega_1(q_2,q_4)\frac{\ln\left(\frac{\mu_1\mu_3}{\mu_2p\prime_-p_+}
 \right)+i\pi}{\mu_1\mu_3+\mu_2p\prime_-p_+}]v(p)\ea
As to the unstable part $M_u$ it turns out to be of the order
$(Z_1Z_2\alpha^2)^3$, i.e. a higher degree in fine structure
constant than (15) and has to be exactly cancelled when one
considers the next terms of the Watson series.
\section{Conclusions}
The problem of taking account of the Coulomb corrections in the
process of lepton pair production in the Coulomb field of two
highly relativistic nuclei turns out to be much more complicated
than seemed decade ago.The hope that due to Lorentz contraction of
the Coulomb potential at ultrarelativistic energies one can obtain
a simple expression for the amplitude was not come true.The
investigation in perturbation theory~\cite{G4} has shown that the
terms of higher order in fine structure constant become more and
more complex and bulky and one needs some general algorithm to
construct them.The receipt proposed in ~\cite{G4} leads to the
infrared unstable result  and thus cannot be considered as a final one.\\
We developed a new approach to this problem using the Watson
series and solving the well-known operator equations for
amplitudes relevant to lepton scattering in the  Coulomb field of
relativistic nuclei and proposed the prescription allowing one
 to construct the amplitude of  process (1) to any order in
$Z_1Z_2\alpha$.We show that in the case of a finite number of
photon exchanges between the lepton pair and one of the ions the
amplitude does not depend on the regularization parameter relevant to this ion.\\
 We would like to thank E.A.Kuraev for valuable discussions and
collaboration on this subject.

\newpage
\renewcommand{\textfraction}{0}
\begin{figure}[ht]
\begin{center}
\includegraphics[scale=1.]{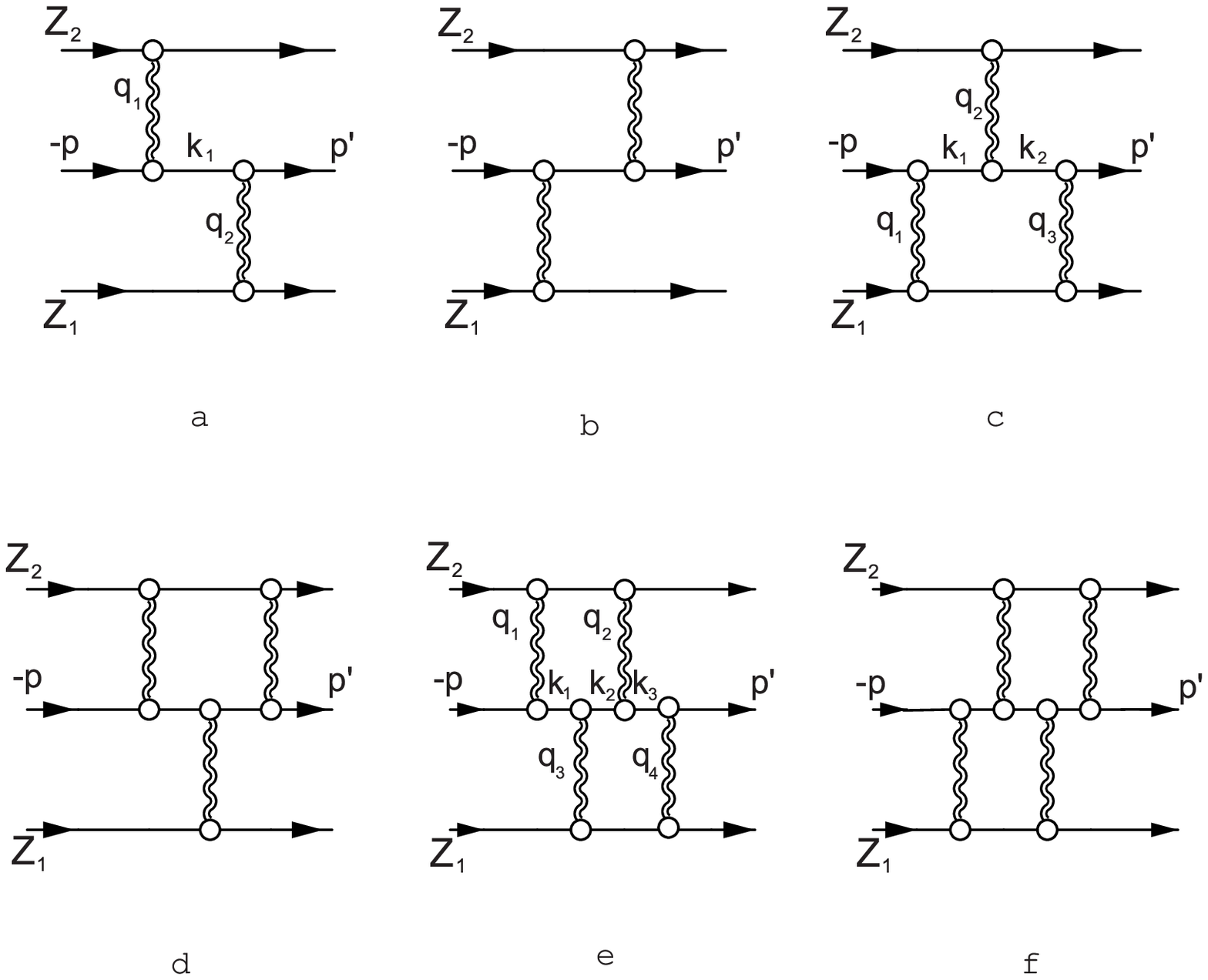}
\caption{The diagrams relevant to first six terms of Watson
series}
\end{center}
\end{figure}
\end{document}